\documentclass[aps,pra,twocolumn,showpacs,superscriptaddress]{revtex4} 
\usepackage{graphicx}
\usepackage{amsmath}
\usepackage{amssymb}
\usepackage{epstopdf}
\usepackage{amsfonts}
\usepackage{dcolumn}
\begin{document}
\title{Electric-field induced
helium-helium resonances}
\author{Q. Guan}
\address{Homer L. Dodge Department of Physics and Astronomy,
  The University of Oklahoma,
  440 W. Brooks Street,
  Norman,
Oklahoma 73019, USA}
\author{D. Blume}
\address{Homer L. Dodge Department of Physics and Astronomy,
  The University of Oklahoma,
  440 W. Brooks Street,
  Norman,
Oklahoma 73019, USA}
\date{\today}

\begin{abstract}
The tunability of the helium-helium interaction through
an external electric field is investigated.
For a static external field, electric-field induced resonances
and associated
electric-field induced bound states
are calculated
for the $^4$He-$^4$He, $^3$He-$^4$He, and $^3$He-$^3$He systems.
Qualitative 
agreement 
is found 
with the literature 
for
the $^3$He-$^4$He 
and
$^3$He-$^3$He systems
[E. Nielsen, D. V. Fedorov,
and A. S. Jensen, Phys. Rev. Lett. {\bf{82}}, 2844 (1999)].
The implications of
the predicted
electric-field induced
resonances for 
$^4$He-$^4$He on the wave packet
dynamics, initiated by intense laser pulses, are investigated.
Our results are expected to guide next generation
experiments.
\end{abstract}
\maketitle

\section{Introduction}
\label{sec_introduction}
The helium atom is a chemically inert rare gas atom.
Helium has two naturally occuring isotopes:
$^4$He, a composite boson, and $^3$He, a composite
fermion.
Whether or not these isotopes form diatomic molecules
was debated for a long time in the 
literature.
It is now agreed upon that the $^4$He-$^4$He system
supports a single rotationless bound state with 
an extremely small binding energy of about
$1.3$~mK~\cite{schoellkopf1994,tang1995,janzen1995,luo1996,schoellkopf1996,grisenti2000,zeller2016}.
Neither the $^3$He-$^4$He  nor the
$^3$He-$^3$He system support, in the absence of
external fields, molecular bound states.

The extremely small binding energy of the $^4$He-$^4$He dimer is associated
with a large positive  $s$-wave scattering length.
The $^3$He-$^4$He system, in contrast, is characterized by
a negative and large, in magnitude, $s$-wave scattering length.
Motivated by the tunability of many of the alkali dimers
through the application of an external magnetic field in the
vicinity of a Fano-Feshbach resonance~\cite{chin2010}, 
one may ask
if the helium-helium interaction can be tuned as well,
with the external magnetic field replaced by an external electric
field.
If such a tunability existed, this would open up
many new research directions related to the study of extremely 
weakly-bound molecular states for a system that is amenable
to {\em{ab initio}} calculations.
Indeed, 
Ref.~\cite{nielsen1999} pointed out the tunability of the
$^3$He-$^4$He and $^3$He-$^3$He systems by a static
external electric field. Moreover,
Ref.~\cite{nielsen1999} explored the consequences of this tunability
for the three-body sector in the context of Efimov 
physics~\cite{efimov70,braaten2006,naidon2016}.
The tunability of the $^4$He-$^4$He system by a static
electric field 
and by laser pulses
strong enough to involve electronically excited potential
curves 
was very recently pointed out in Ref.~\cite{QiWei}.

Working in the opposite regime of short
laser pulses,
a recent molecular beam experiment~\cite{kunitski2018}
demonstrated that a short 310fs laser pulse with
an intensity of a few times  $10^{14}$W/cm$^2$
can induce dissociative wave packet dynamics 
of the $^4$He-$^4$He dimer, including interferences
between the $l=0$ and $l=2$ partial wave channels.
Here, $l$ denotes the orbital angular
momentum quantum number.
While Ref.~\cite{kunitski2018} provided no
evidence for the existence of electric field-induced
bound states or hybridized states
such as those predicted in Ref.~\cite{friedrich1998}, 
the experimental results
clearly show
that the laser-molecule
coupling is strong enough to trigger
measurable changes such as a clean
alignment signal. Moreover, the excellent agreement between
the experimental and theoretical results 
in Ref.~\cite{kunitski2018} suggests that the 
laser-molecule interaction, which included the 
lowest Born-Oppenheimer potential curve and 
assumed inertness of the electronic
degrees of freedom, provides a reliable
description, at least in the short-pulse regime
for the intensities considered.

The present theoretical work considers 
laser pulses that are
longer than those utilized in Ref.~\cite{kunitski2018}.
As a first
exploration, our theoretical framework neglects, 
as in Ref.~\cite{kunitski2018},
the electronic degrees of freedom.
It is expected that corrections due to the electronic
motion (see, e.g., Ref.~\cite{becker})
need to be accounted for in follow-up work.
One of the goals is to explore under which conditions the electric-field induced
resonances of the helium-helium
systems, first investigated
in Ref.~\cite{nielsen1999}
for the $^3$He-$^4$He and $^3$He-$^3$He systems and 
for the $^4$He-$^4$He system in Ref.~\cite{QiWei},
can be observed experimentally in time-dependent
set-ups.
To interpret the dynamic wave packet studies, the static 
field case is revisited and 
some quantitative 
discrepancies with the 
literature~\cite{nielsen1999},
which we have no explanation for,
are pointed out. 
To observe the electric-field induced resonances,
the associated bound states have to be populated with sufficiently
high probability and some signature that this 
has been achieved needs to be recorded.
Our time-dependent calculations show, owing to the
extremely floppy and highly quantum mechanical nature
of the helium dimers, that revival dynamics
competes with dissociative dynamics.
Probing this intricate dynamics experimentally is expected to be
possible but quite challenging 
due to the 
need of realizing long, intense laser pulses.

The remainder of this 
article is organized as follows. Section~\ref{sec_background}
introduces the system Hamiltonian and relevant
theoretical background.
Sections~\ref{sec_results1} and \ref{sec_results2}
present our results for a static external field and 
a time-dependent external field, respectively.
Last, Sec.~\ref{sec_conclusion} concludes.

\section{System Hamiltonian and Theoretical Background}
\label{sec_background}
This section describes the theoretical framework
employed to investigate the tunability of the 
effective helium-helium
interaction strength.
Section~\ref{sec_ham}
introduces the system Hamiltonian.
The determination of the scattering and bound states of the static
Hamiltonian are 
discussed in Secs.~\ref{sec_scattering} and \ref{sec_bound}.
Last, Sec.~\ref{sec_dynamics} summarizes how the wave packet 
propagation is done when the Hamiltonian is time-dependent.

\subsection{System Hamiltonian}
\label{sec_ham}
We consider two helium atoms,
either two $^4$He atoms, a $^3$He-$^4$He pair, or two $^3$He atoms, 
with reduced
mass $\mu$ interacting through the spherically-symmetric
state-of-the-art ``electronic ground state'' potential 
$V_{\text{2b}}(r)$ from Ref.~\cite{cencek2012}, where 
$\vec{r}$ denotes the internuclear distance vector
and $r$ is equal to $|\vec{r}|$.
Due to the adiabatic beyond Born-Oppenheimer correction 
term~\cite{cencek2012},
the interaction potentials for $^4$He-$^4$He, $^3$He-$^4$He,
and $^3$He-$^3$He
are slightly different.

Throughout we assume that the electric field of
the laser is oriented along the $z$-axis.
Moreover, we assume that the oscillations of the electric field are 
so fast that they can be integrated over.
With these assumptions, the time-dependent laser-molecule interaction
$V_{\text{lm}}(r,\theta,t)$ reads~\cite{friedrich1995}
\begin{eqnarray}
  V_{\text{lm}}(r , \theta,t)= -\frac{1}{2} |\epsilon(t)|^2
  \left[ \alpha_{\parallel}(r) \cos^2 \theta + \alpha_{\perp}(r) \sin^2 \theta \right],
  \end{eqnarray}
where $\theta$ denotes the angle between the $z$-axis and the internuclear 
distance vector
$\vec{r}$ (in spherical coordinates, this is the azimuthal angle),
$\epsilon(t)$ characterizes the shape of the laser pulse, and
$\alpha_{\perp}(r)$ 
and $\alpha_{\parallel}(r)$ denote the polarizabilities 
perpendicular and parallel to the molecular axis. The 
difference between these polarizabilities 
is responsible for the intriguing
dynamics discussed in Sec.~\ref{sec_results2}. 

\begin{figure}
  \centering
  \vspace*{0.5in}
\includegraphics[angle=0,width=0.4\textwidth]{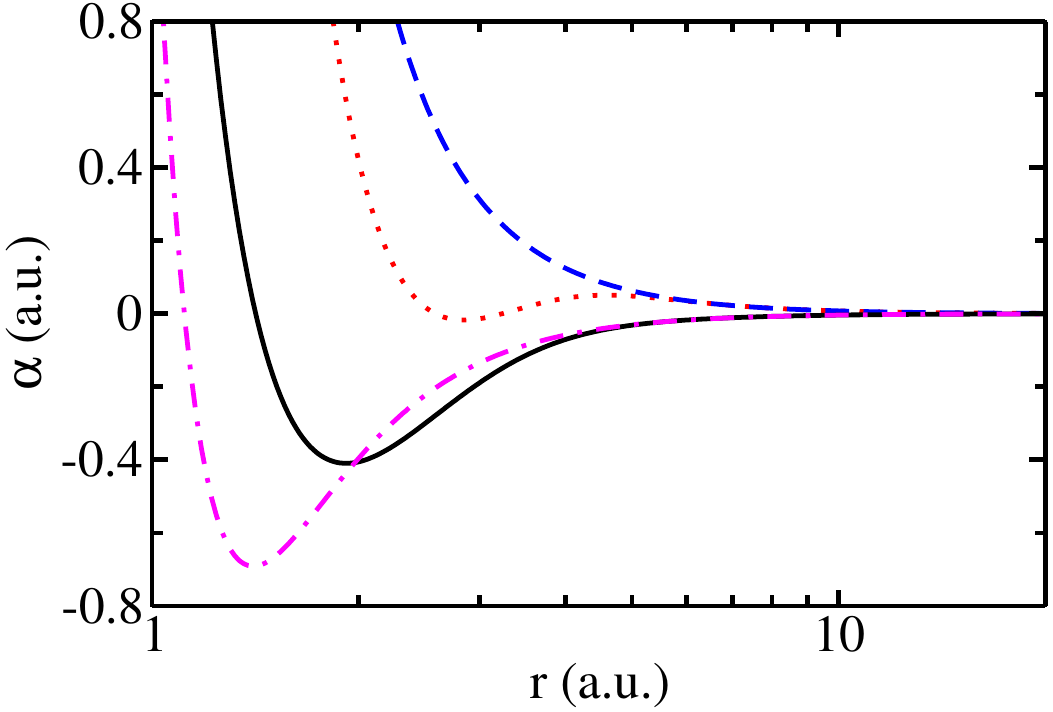}
\vspace*{0.5in}
\caption{(Color online)
Polarizabilities as a function of the internuclear distance $r$
  (note the logarithmic scale of the horizontal axis).
  The solid and dotted lines show $\alpha_{\perp}$ and $\alpha_{\parallel}$
  using the {\em{ab initio}} data from Ref.~\cite{cencek2011}.
  The dash-dotted and dashed lines show $\alpha_{\perp}$ and $\alpha_{\parallel}$
  using the analytical polarizabilities [see Eqs.~(\ref{eq_alpha_perp})
    and (\ref{eq_alpha_parallel})].
  The constant contribution of $2 \alpha_0$
  is not included in the plots.
   }\label{fig_polarizability}
\end{figure}

Following the pioneering work of Buckingham and 
Watts~\cite{buckingham1973}, analytic
expressions for $\alpha_{\perp}(r)$ and $\alpha_{\parallel}(r)$
read
    \begin{eqnarray}
\label{eq_alpha_perp}
    \alpha_{\perp}(r) = 2 \alpha_0 - 
\frac{2 \alpha_0^2}{4 \pi {\cal{E}}_0 \,  r^3} + 
\frac{2 \alpha_0^3}{(4 \pi {\cal{E}}_0)^2 \, r^6}
  \end{eqnarray}
and  
\begin{eqnarray}
\label{eq_alpha_parallel}
    \alpha_{\parallel}(r) = 2 \alpha_0 + 
\frac{4 \alpha_0^2}{4 \pi {\cal{E}}_0 \, r^3} + 
\frac{8 \alpha_0^3}{(4 \pi {\cal{E}}_0)^2 \, r^6},
  \end{eqnarray}
where $\alpha_0$ denotes the atomic polarizability,
$\alpha_0=1.383a.u.$ (``$a.u.$'' stands for ``atomic units''),
and ${\cal{E}}_0$ the permittivity ($4 \pi {\cal{E}}_0 = 1 a.u.$;
note that the symbols $\epsilon$ and ${\cal{E}}_0$ refer to different
physical quantities).
To discuss the physics, we rewrite $V_{\text{lm}}$
(in doing so, we drop the $r$-independent terms, which only
contribute an energy shift and/or an overall phase),
\begin{eqnarray}
  \label{eq_vlm_physics}
  V_{\text{lm}}(r,\theta,t)&=&
   \frac{|\epsilon(t)|^2 \alpha_0^2}{4 \pi {\cal{E}}_0}
  \Bigg[
    -\frac{2\alpha_0}{(4 \pi {\cal{E}}_0)\, r^6} + 
    \nonumber \\
    &&
    \alpha_0 \frac{1-3 \cos ^2 \theta}{(4 \pi {\cal{E}}_0) \, r^6}
    +
    \frac{1-3 \cos ^2 \theta}{r^3}
    \Bigg].
\end{eqnarray}
The first term in square brackets
shows that the laser-molecule interaction
increases the $C_6$ van der Waals coefficient of the helium-helium
potential.
The second term in square brackets shows that the laser-molecule interaction
introduces an angle-dependence into the $C_6$ coefficient.
Finally, the third term in square brackets corresponds to the interaction
between two point dipoles, yielding a repulsive interaction energy
for a side-by-side configuration and an attractive interaction energy
for a head-to-tail configuration.
These analytic expressions agree well with 
the state-of-the-art 
{\em{ab initio}} results from Ref.~\cite{cencek2011} in the large $r$
region but not in the small $r$ region 
(see Fig.~\ref{fig_polarizability}).

We find that 
the analytic expressions and
the {\em{ab initio}} parametrization yield 
predictions
that differ quantitatively but not qualitatively.
The analytic
polarizability model,
for example, supports field-induced bound states for somewhat
smaller field strengths than the {\em{ab initio}} parametrization.
Similarly, the dynamical results 
presented in Sec.~\ref{sec_results2} are
dominated by 
the polarizabilities around $4a.u.$ to $10a.u.$ for which
the two sets of polarizabilities agree quite well.
Since the results for the two models agree qualitatively,
the majority
of the 
results presented in this 
work employs the polarization model from Ref.~\cite{cencek2011}.

Since
the Hamiltonian $H$ is in
our set-up
independent of the polar angle $\phi$,
the projection quantum number $m_l$, which is associated with the
$z$-component of the orbital angular momentum operator $\vec{l}$,
is a good quantum number.
We restrict ourselves to the $m_l=0$ channel in this work.
Combining the interaction terms, the time-dependent  Hamiltonian $H$,
written in spherical coordinates,
reads
\begin{eqnarray}
H = -\frac{\hbar^2}{2 \mu} \Bigg[
\frac{1}{r^2} \frac{\partial}{\partial r} \left( r^2 \frac{\partial}{\partial r}
\right)+ \nonumber \\
\frac{1}{r^2 \sin \theta} \frac{\partial}{\partial \theta}
\left(
\sin \theta \frac{\partial}{\partial \theta} \right) \Bigg]
+
V_{\text{2b}}(r) +
V_{\text{lm}}(r,\theta,t).
\end{eqnarray}

We consider two
parametrizations of $\epsilon(t)$:
\begin{enumerate}
\item Static field with 
$\epsilon(t)=\epsilon_{0,\text{S}}$, 
where
$\epsilon_{0,\text{S}}$ is a constant. Even though some of the field strengths 
$\epsilon_{0,\text{S}}$ considered
in this work can only be realized for
a relatively short time  with present day technology, the results
for the static field provide a useful framework for understanding the
results for time-dependent pulses.
\item A ``stretched'' Gaussian pulse with $\epsilon(t)=\epsilon_{\text{SG}}(t)$,
\begin{eqnarray}
\epsilon_{\text{SG}}(t)=
\left\{
\begin{array}{lll}
\epsilon_{\text{G}}(t) & \mbox{for} & t \le 0 \\
 \epsilon_{0,\text{G}} & \mbox{for} & 0 < t < t_{\text{hold}} \\
\epsilon_{\text{G}}(t-t_{\text{hold}}) & \mbox{for} & t_{\text{hold}} \le t,
\end{array}
\right.
\end{eqnarray}
where
  \begin{eqnarray}
    \epsilon_{\text{G}}(t) = \epsilon_{0,\text{G}} \exp \left( -2 \ln(2) \frac{t^2}{\tau^2}
    \right) 
  \end{eqnarray}
  with $\ln(2)=0.6931\cdots$.  At $t=0$, the
pulse is maximal for the first time and $\tau$ is the FWHM, which
determines the rise and fall-off of the
Gaussian pulse. 
While the stretched Gaussian pulse shape may not be realizable 
experimentally, the ensuing dynamics is comparatively straightforward 
to interpret
and thus serves as a guide to what might be expected
for stretched pulses with somewhat different profiles.
\end{enumerate}

\subsection{Scattering States}
\label{sec_scattering}
In the absence of the external field ($V_{\text{lm}}=0$), the $s$-wave scattering 
length $a_s$
of the $^4$He-$^4$He system is positive and large
($a_s=170.9a.u.$), signaling the existence of a weakly-bound
molecular $s$-wave state. In fact, this is the only bound state supported
in the field-free case; no rotationally or vibrationally excited states exist.
The $s$-wave scattering length of the $^3$He-$^4$He system, in contrast, is negative
and large in magnitude 
($a_s=-34.2a.u.$), signaling that the system is just short
of supporting a weakly-bound $s$-wave bound state. No deep-lying bound states
are supported.
The $s$-wave scattering length of the $^3$He-$^3$He system
is equal to $-13.73a.u.$ in the absence
of an external electric field;
in this case, the nuclear spins form a singlet, thereby enforcing the
anti-symmetry of the full
wave function under the exchange of two identical $^3$He atoms. 
The magnitude of the
generalized higher partial wave scattering lengths such as 
the $p$-wave scattering volume for the $^3$He-$^4$He and 
$^3$He-$^3$He systems and the 
$d$-wave scattering hypervolume for the $^4$He-$^4$He system are small.

We now include a time-independent laser-molecule Hamiltonian
(parametrization 1.~in Sec.~\ref{sec_ham}), 
which couples different orbital angular momentum
channels. 
For the bosonic $^4$He-$^4$He system, only even-$l$ channels
contribute because the spatial wave function has to be
symmetric under the exchange of the two $^4$He atoms.
For the $^3$He-$^4$He system, in contrast,
no symmetry constraints exist, implying that 
even- and odd-$l$ channels contribute (due to the nature
of the laser-molecule interaction, the even- and odd-$l$
channels are decoupled).
Last, for the fermionic $^3$He-$^3$He system, even-$l$
channels contribute when the nuclear spins form a singlet and odd-$l$ channels
when the nuclear spins form a triplet.
The long-range nature of the laser-molecule interaction
modifies the threshold law
in the non-zero partial wave
channels~\cite{weiner1999,marinescu1998,yi2001,deb2001}.
In particular, since the K-matrix elements $K_{l,l'}(k)$,
\begin{eqnarray}
\label{eq_kmatrix}
K_{l,l'}(k) = \tan \left( \delta_{l,l'}(k) \right),
\end{eqnarray}
are proportional to the
wave vector $k$ as $k$ goes to zero
($k$ is defined in terms of the scattering energy $E$ through
$\sqrt{2 \mu E}/\hbar$),
the zero-energy
scattering length matrix elements $a_{l,l'}$
are defined through
\begin{eqnarray}
\label{eq_ascatt}
a_{l,l'} = \lim_{k \rightarrow 0} \frac{- K_{l,l'}(k)}{k}.
\end{eqnarray}
For short-range interactions
(interactions that fall off faster than 
$1/r^3$ at large internuclear distances), 
the denominator in Eq.~(\ref{eq_ascatt})
reads $k^{l+l'+1}$ instead of $k$~\cite{newton_book}; 
the modification of the power of $k$
reflects the modified threshold behavior.
The threshold laws, Eqs.~(\ref{eq_kmatrix}) and (\ref{eq_ascatt}),
depend crucially on the angle dependence of the
$-r^{-3}$ potential. If the angle dependence was absent, one would not be able to
define an $s$-wave scattering length.

The phase shifts $\delta_{l,l'}(k)$ [see Eq.~(\ref{eq_kmatrix})]
are obtained by matching the inside solution to
the large-$r$, free-particle 
solution, with the 
relative importance of the regular solutions
[the spherical Bessel functions $j_l(kr)$] and the irregular 
solutions
[the Neumann functions $n_l(kr)$] given by the 
tangent of the phase shifts $\delta_{l,l'}(k)$.
The scattering solutions in the presence of a static
external field are thus characterized by 
an, in general,
non-diagonal scattering length
matrix.
Even though the determination of the scattering solutions
requires the entire scattering length matrix,
the emergence of a new zero-energy
bound state that is even (odd) in the relative
coordinate $z$ is accompanied by the $a_{0,0}$ 
($a_{1,1}$) matrix element 
going to infinity~\cite{ticknor2005,kanjilal2008}.

We determine the K-matrix by 
decomposing the
full wave function $\psi(r,\theta)$
into partial waves,
\begin{eqnarray}
\label{eq_expand}
\psi(r,\theta) = \sum_{l'} \frac{u_{l'}(r)}{r} Y_{l',0}(\cos \theta),
\end{eqnarray}
where the sum over $l'$ includes all angular momentum
values allowed by symmetry and where the spherical harmonics
$Y_{l',m_l'}$ are independent of $\phi$ since $m_l'$ is assumed to be
zero throughout.
Inserting Eq.~(\ref{eq_expand}) into the 
Schr\"odinger equation $H \psi = E \psi$
and projecting onto the $Y_{l,0}^*$ states, 
we obtain a 
set of coupled 
differential equations for the 
radial components $u_{l}(r)$,
\begin{eqnarray}
 \left(
-\frac{\hbar^2}{2 \mu} \frac{\partial^2 }{\partial r^2} +
V_{\text{2b}}(r) \right) u_{l}(r)
- \nonumber \\
\frac{1}{2}|\epsilon_{0,\text{S}}|^2 \sum_{l'} W_{l,l'}(r) u_{l'}(r)
= E u_{l}(r).
\end{eqnarray}
An explicit
expression for the coupling matrix elements $W_{l,l'}(r)$,
which arise from integrating over the angular 
degrees of freedom,
is given in Appendix~\ref{appendix_coupling}.
Equation~(\ref{eq_coupling}) shows that the laser-molecule
interaction $V_{\text{lm}}$ couples only channels with the same
$l$ or channels whose indices differ by two.
Imposing that the $u_{l}(r)$ vanish at small $r$,
the logarithmic derivative 
matrix is propagated using the Johnson algorithm~\cite{johnson}
with adjustable step size. 
Matching the large-$r$ solution
to the asymptotic
free-particle solution, the K-matrix is extracted.
We find that a
scattering energy of $10^{-12}a.u.$
approximates the zero-energy limit accurately; for this energy,
we choose the large-$r$ matching point to be $10^6 a.u.$.
We find that the inclusion of about $8$ 
even and/or 8 odd partial wave channels 
yields converged results for the field strengths
considered in this work.

\subsection{Bound States}
\label{sec_bound}
In addition to the scattering states, we calculate the
bound states of the helium-helium systems in a static
external field. 
Since a time-dependent external field can, at each time, be thought of as
being static,
the solutions for the static Hamiltonian
provide guidance for interpreting our time-dependent results.
In the extreme case of an adiabatically changing external field,
the full dynamics can be readily extracted from the
static results by, e.g., performing a Landau-Zener analysis.

To determine the bound state spectrum, we express the
eigen states $\psi(r,\theta)$ in terms of a B-spline basis
using non-linear grids in $r$ and $\theta$.
The largest $r$ is adjusted 
so that the most weakly-bound state is fully covered by the 
numerical grid.
For $^4$He-$^4$He,
we calculate 
eigen states that are
even 
in the relative coordinate $z$. For the $^3$He-$^3$He and $^3$He-$^4$He systems,
both even and odd states in $z$ are considered.
In the case of $^3$He-$^3$He, the even and odd partial waves must be combined
with singlet and triplet nuclear spin states, respectively.
Even though the bound states cannot 
be labeled by a single $l$ quantum number
due to the $\theta$-dependence of $V_{\text{lm}}$,
the weakly-bound states are typically dominated 
by a single partial wave.
The dominant character can be obtained by projecting the
eigen states onto different $l$ channels.

\subsection{Dynamics}
\label{sec_dynamics}

If the laser-molecule interaction is
time dependent, we have to solve the time-dependent
Schr\"odinger equation for a given initial state
$\Psi(r,\theta,t=-\infty)$.
In practice, the initial state is prepared 
at a time where the laser-molecule interaction can be neglected,
i.e., at a time much smaller than 
$0$.

To solve the time-dependent Schr\"odinger equation,
we decompose the wave packet $\Psi(r,\theta,t)$, similar
to what we did in Sec.~\ref{sec_scattering} to obtain the time-independent
scattering states,
into partial wave components,
\begin{eqnarray}
\label{eq_expand_time}
\Psi(r,\theta,t) = \sum_{l'} \frac{U_{l'}(r,t)}{r} Y_{l',0}(\cos \theta).
\end{eqnarray}
Inserting Eq.~(\ref{eq_expand_time}) into the time-dependent
Schr\"odinger equation
$\imath \hbar \partial \Psi/ \partial t = H \Psi$,
we obtain a 
set of coupled 
time-dependent 
equations
for the radial components $U_l(r,t)$,
\begin{eqnarray}
 \left(
-\frac{\hbar^2}{2 \mu} \frac{\partial^2 }{\partial r^2} +
V_{\text{2b}}(r) \right) U_{l}(r,t)
- \nonumber \\
\frac{1}{2} | \epsilon(t)|^2 \sum_{l'} W_{l,l'}(r) U_{l'}(r,t)=
\imath \hbar \frac{\partial U_l(r,t)}{\partial t},
\end{eqnarray}
where the coupling elements $W_{l,l'}(r)$
are given in Eq.~(\ref{eq_coupling}).

To solve the coupled set of time-dependent radial equations,
we discretize the $r$ coordinate (we typically use about 32,000
points) and propagate the $U_l(r,t)$
by expanding the radial propagator in terms of Chebychev 
polynomials~\cite{kosloff}.
The time step $\Delta t$
is chosen such that $|\epsilon(t)|^2$ 
can be considered, to a very good approximation, 
as time independent during each time step. 
We use $\Delta t \approx 0.24~\text{fs}$
to
$0.60~\text{fs}$ and about
$30$ terms in the 
expansion
into Chebychev
polynomials. For the pulses considered, accounting
for about
eight partial wave channels yields converged results.

\section{Time-Independent Field Strength}
\label{sec_results1}

\begin{figure}
  \centering
  \vspace*{0.5in}
\includegraphics[angle=0,width=0.4\textwidth]{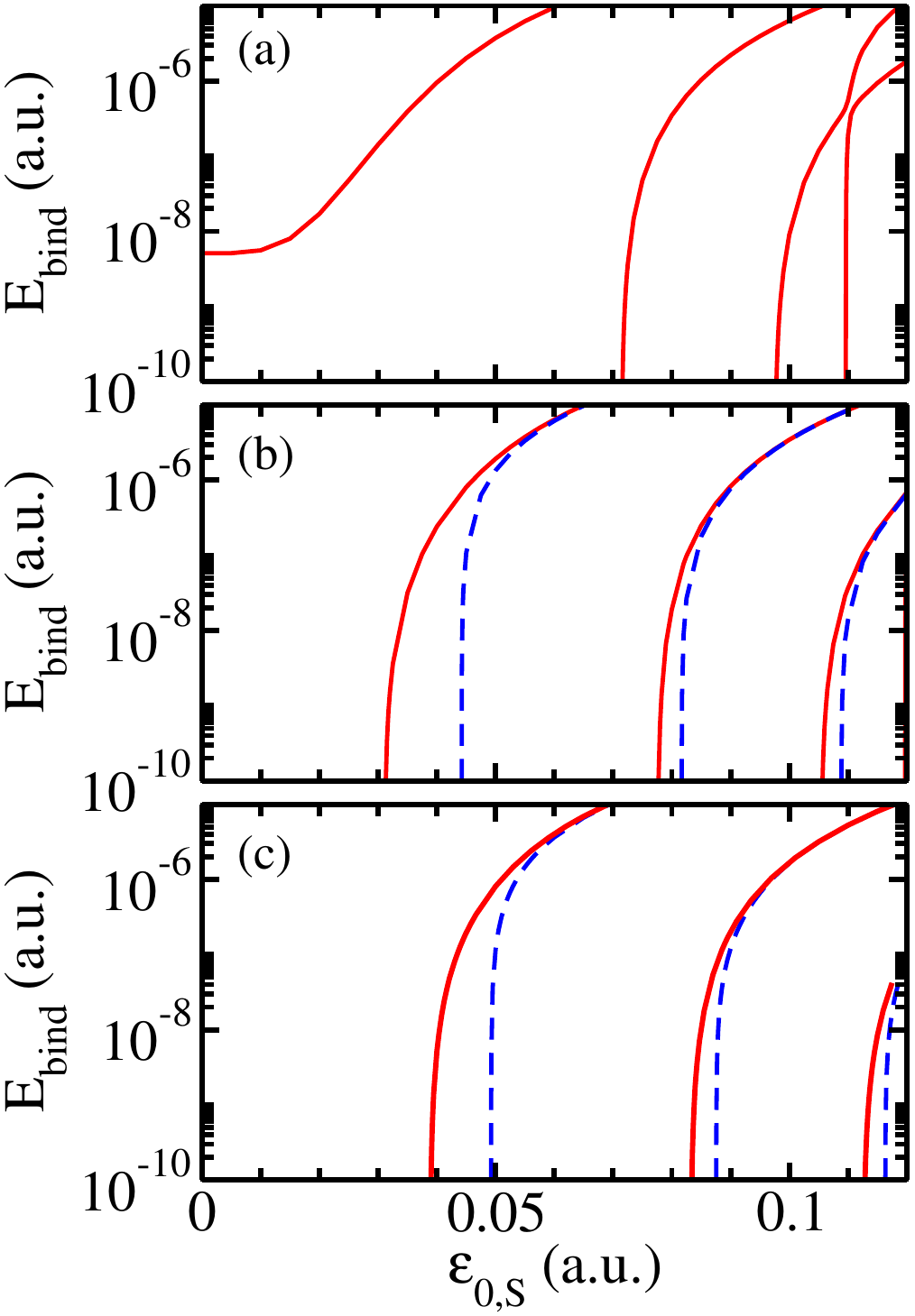}
\vspace*{0.5in}
\caption{(color online)
Binding energy $E_{\rm{bind}}$ for
(a) $^4$He-$^4$He,
(b) $^3$He-$^4$He,
and
(c) $^3$He-$^3$He
as a function of the field strength $\epsilon_{0,\text{S}}$.
The solid lines show the 
binding energy of states that are even in
$z$ while the dashed lines show the 
binding energy of states that are 
odd in
$z$.
Note that the binding energy is shown on a logarithmic
scale that covers five orders of magnitude.
 }\label{fig_energy}
\end{figure}

This section discusses the characteristics of the helium-helium
systems in the presence of a static external field 
(parametrization 1.~in Sec.~\ref{sec_ham}).
Figure~\ref{fig_energy} shows the dimer binding energy
$E_{\rm{bind}}$
for (a) $^4$He-$^4$He,
(b) $^3$He-$^4$He, and
(c) $^3$He-$^3$He
as a function of the field strength $\epsilon_{0,\text{S}}$.
The binding energy associated with states that are even in $z$ is shown
by solid lines and that associated with states that are odd in
$z$ is shown by dashed lines.
The $^4$He-$^4$He system supports new
$s$-wave ($l=0$) dominated
bound
states for  field strengths larger than about  $\epsilon_{0,\text{S}}=0.0715a.u.$
and larger than about $\epsilon_{0,\text{S}}=0.0976a.u.$.
No evidence for the existence of these states is
reported in Ref.~\cite{QiWei}.
For field strengths larger than about
$\epsilon_{0,\text{S}}=0.10962a.u.$, a new 
bound state with appreciable $d$-wave admixture
is being supported
[also notice the related avoided crossing between the 
$s$-wave dominated and $d$-wave dominated states at
$(\epsilon_{0,\text{S}},E_{\rm{bind}})
\approx
(0.11a.u.,5 \times 10^{-7}a.u.)$]. 
Owing to the orbital angular momentum
barrier, this bound state acquires an
appreciable binding energy over a fairly small variation
of
the field strength $\epsilon_{0,\text{S}}$.
Reference~\cite{QiWei} refers to this $d$-wave dominated state
as a ``pendular state''.

The $^3$He-$^4$He system 
[see Fig.~\ref{fig_energy}(b)]
supports its first $s$-wave dominated bound state
for field strengths larger than about $\epsilon_{0,\text{S}}=0.0311a.u.$,
a second $s$-wave 
dominated bound state for field strengths larger than about 
$\epsilon_{0,\text{S}}=0.0776a.u.$, and
a third $s$-wave 
dominated bound state for field strengths larger than about 
$\epsilon_{0,\text{S}}=0.1054a.u.$.  
Owing to the
smaller reduced mass, the latter two field strenths are a bit
larger than the critical field strengths for the $^4$He-$^4$He system.
The first field-induced bound state,
which first appears at $\epsilon_{0,\text{S}}=0.0311a.u.$,
has no analog in the $^4$He-$^4$He system since this
system already supports a weakly bound state in the absence of
an external electric field.
The $^3$He-$^4$He system additionally supports bound states
that are odd in the relative coordinate $z$ [see the dashed
lines in Fig.~\ref{fig_energy}(b)]. 
Interestingly, these odd-$z$ bound states first appear
at field strengths that are just a bit larger than the
field strengths at which the even-$z$ bound
states first appear. 
As the binding energy increases, the energy difference,
normalized by the binding energy itself, between
the pairs of even-$z$ and odd-$z$ states decreases.
This is not unlike the tunneling splitting in a
double-well potential, where the tunneling is much smaller 
for deep-lying states than for states that lie near
or above the barrier.

Last, 
the $^3$He-$^3$He system first supports even-$z$ bound states at field
strengths
larger than about 
$\epsilon_{0,\text{S}}=0.0388a.u.$, larger than about $0.0832a.u.$, and 
larger than about $0.1126a.u.$, respectively. 
Odd-$z$
bound states are first supported at field strengths larger than
about $\epsilon_{0,\text{S}}=0.0492294a.u.$, larger than about $\epsilon_{0,\text{S}}=0.0875679a.u.$,
and larger than about $\epsilon_{0,\text{S}}=0.116342a.u.$.
Owing to the
smaller reduced mass, these field strenths are a bit
larger than the 
corresponding critical 
field strengths for the $^3$He-$^4$He system.

\begin{figure}
\centering
\includegraphics[angle=0,width=0.4\textwidth]{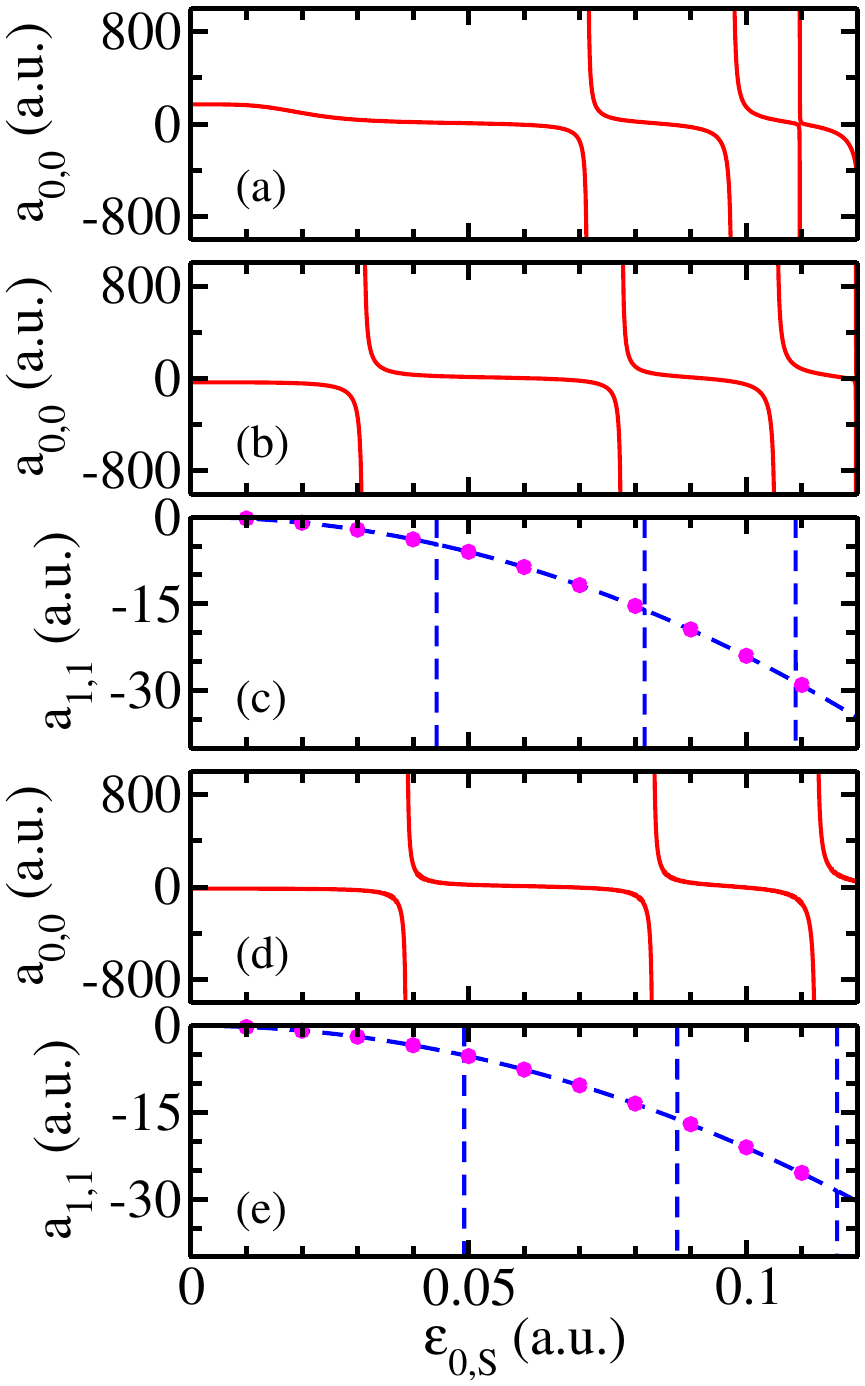}
\vspace*{0.5in}
\caption{(Color online)
Scattering lengths for (a) $^4$He-$^4$He,
(b) and (c) $^3$He-$^4$He,
and
(d) and (e) $^3$He-$^3$He
as a function of the field strength $\epsilon_{0,\text{S}}$.
The solid lines in (a), (b), and (d) show the scattering length $a_{0,0}$
while the dashed lines in (c) and (e)
show the scattering length $a_{1,1}$.
The narrow resonance at $\epsilon_{0,\text{S}} \approx 0.11 a.u.$
in (a) has notable $d$-wave admixture.
The resonances in the $(l,l')=(1,1)$ channel
[see panels (c) and (e)] are extremely narrow.
The solid circles in panels (c) and (e)
show $a_{1,1}$ as predicted by the Born approximation;
the Born approximation reproduces the ``background
value'' very well but does not capture the resonances
(the scattering lengths in the Born approximation are
directly proportional to $-|\epsilon_{0,\text{S}}|^2$).
 }\label{fig_ascatt}
\end{figure}

\begin{figure}
\centering
\includegraphics[angle=0,width=0.4\textwidth]{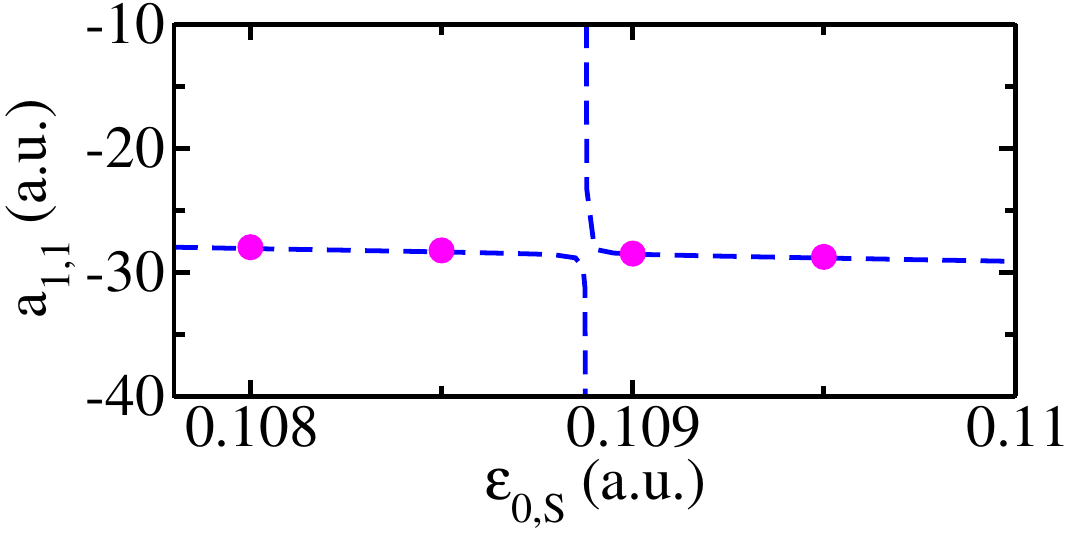}
\vspace*{0.0in}
\caption{(Color online)
Enlargement of 
the scattering length $a_{1,1}$ for $^3$He-$^4$He
in the vicinity of a resonance.
The data and symbol/line styles are the same
as in Fig.~\ref{fig_ascatt}(c).}
\label{fig_scatt_blowup}
\end{figure}

As mentioned in Sec.~\ref{sec_scattering},
the emergence of a new even-$z$
bound state is accompanied by a diverging 
$a_{0,0}$
and
the emergence of a new odd-$z$
bound state by a diverging 
$a_{1,1}$.
Solid lines
in Fig.~\ref{fig_ascatt} show the scattering length matrix
element
$a_{0,0}$ as a function of the field strength $\epsilon_{0,\text{S}}$
for (a) $^4$He-$^4$He, (b) $^3$He-$^4$He, and (d) $^3$He-$^3$He
while dashed  lines
show the scattering length matrix
element
$a_{1,1}$ 
for (c) $^3$He-$^4$He and (e) $^3$He-$^3$He.
Comparison with Fig.~\ref{fig_energy} shows that
$a_{0,0}$ and $a_{1,1}$ go through infinity at the field strengths
at which new, respectively, even-$z$ and odd-$z$
bound states are first being supported.
We checked that the generalized scattering lengths $a_{l,l'}$,
except for $a_{0,0}$,
are well described---as they should
be
for potentials that are purely dipolar at large
internuclear distances~\cite{yi2001}---by
the Born approximation for field strengths
where resonances are absent.
This is illustrated in Figs.~\ref{fig_ascatt}(c) and
\ref{fig_ascatt}(e), where the Born approximation results
(solid circles) reproduce the full 
coupled-channel 
calculations (dashed lines) reliably.
In the Born approximation,
$a_{1,1}$ is given by
$-2 \mu |\epsilon_{0,\text{S}}|^2 |\alpha_0|^2 /[5 (4 \pi {\cal{E}}_0)^2  \hbar^2]$~\cite{kanjilal2007}.
Figure~\ref{fig_scatt_blowup}
shows an enlargement 
of the scattering length matrix element $a_{1,1}$ for $^3$He-$^4$He
near the third resonance shown in Fig.~\ref{fig_ascatt}(c), i.e.,
near $\epsilon_{0,\text{S}} \approx 0.10885a.u.$. The Born approximation
values do not capture the resonance; instead, they
continue to change quadratically with 
$\epsilon_{0,\text{S}}$ across the resonance.

The calculations presented thus far employ the
{\em{ab initio}} polarization model from Ref.~\cite{cencek2011}.
If we use the simpler analytical polarization model
[Eqs.~(\ref{eq_alpha_perp}) and (\ref{eq_alpha_parallel})],
the first electric-field induced resonance 
for $^4$He-$^4$He occurs at $\epsilon_{0,\text{S}}=0.0699a.u.$
instead of at $\epsilon_{0,\text{S}}=0.0715a.u.$
and
the first electric-field induced resonance
for $^3$He-$^4$He occurs at $\epsilon_{0,\text{S}}=0.0304a.u.$
instead of at $\epsilon_{0,\text{S}}=0.0311a.u.$.
The deviation between the results for the two different
polarization models increases with increasing field strength. 

Our results for the scattering properties of the 
$^3$He-$^4$He and
$^3$He-$^3$He systems in the presence of a 
static field disagree quantitatively
with those presented in Ref.~\cite{nielsen1999, footnote}.
Repeating the calculations for the
interactions employed in Ref.~\cite{nielsen1999}
(i.e., using the LM2M2 potential~\cite{aziz1991} 
and the analytic polarizability model), 
we find 
that the first bound state
for $^3$He-$^4$He is supported for $\epsilon_{0,\text{S}}=0.0305a.u.$
as opposed to $0.053a.u.$ as reported in Ref.~\cite{nielsen1999}
and the first $s$-wave dominated bound state
for $^3$He-$^3$He first appears at
$\epsilon_{0,\text{S}}=0.0382a.u.$ as opposed to
$\epsilon_{0,\text{S}}=0.067a.u.$ as reported in Ref.~\cite{nielsen1999}.
We have no insight
into what might be the reason for the
discrepancies.

The results presented so far indicate that the
helium-helium interaction strength can be
varied through the application of a static external field.
While the field strengths required are attainable
with present-day technology, they can only be realized
for a relatively short time. For the field to be considered 
truly static, the pulse duration has to be longer than the 
internal or characteristic  time scale of the helium-helium
system. If
we convert the $^4$He-$^4$He binding energy in the absence of
an external field,
we find a time
scale of about 
$29.71$~ns. Clearly, the realization of such temporally extended, high-intensity laser
pulses is presently out of reach.
Alternatively, the minimal energy of the
He-He interaction potential corresponds to a time scale of
about $4.364$~ps.
This time scale estimate looks much more promising from an experimental
point of view.
Alternatively, we can estimate the time scale associated 
with the field-induced bound states.
Energies of $10^{-10}a.u.$, $10^{-8}a.u.$, and $10^{-6}a.u.$
correspond to time scales of 
about 
$1,520$~ns, $15.20$~ns, and $0.1520$~ns.
If one were to populate the new bound state and if the system could be
held at a particular field strength longer than the time
given above, one should be able to see revival signatures
corresponding to the above time scale in the dynamical
evolution of an appropriately chosen observable.
While challenging, realizing such a scenario experimentally
does not
seem entirely out of reach.
Ultimately, one has to analyze the full dynamics to 
see which pulse shapes and lengths yield observable signatures
of the electric-field induced tunability of the helium-helium 
interaction strength. Exploratory
calculations along these lines are presented in 
the next section.
While earlier work~\cite{bruch2000} 
employed a perturbative framework
to address this question, we employ a full coupled-channel
treatment. Our
calculations employ 
peak electric field strengths 
of $\epsilon_{0,\text{G}}=0.0843949a.u.$
and $0.11a.u.$
(corresponding to $2.5\times 10^{14}$W/cm$^{2}$ and
$4.247\times 10^{14}$W/cm$^{2}$, respectively).
These field strengths are significiantly lower
than those employed in ``realm II'' of Ref.~\cite{QiWei}.

\section{Time-Dependent Field Strength}
\label{sec_results2}

This section summarizes our results for
the stretched Gaussian pulse (parameterization 2. in Sec.~\ref{sec_ham}).
Our studies are motivated
by two questions:
What, if any, are the signatures 
of the field-induced resonances discussed in Sec.~\ref{sec_results1}
that could be measured experimentally in pump-probe experiments?
Do the field-induced resonances 
lead to revival dynamics, somewhat reminiscent of what
has been observed in pump-probe experiments for stiff, rigid rotor-like
diatomic molecules~\cite{RMP_alignment,review_lemeshko}?
To address these questions, we focus on the
$^4$He-$^4$He system.
We assume that the dimer is prepared in 
the absence of an external field
in its $l=0$ ground state,
as is being done in molecular beam 
experiments~\cite{schoellkopf1994,schoellkopf1996,grisenti2000,zeller2016,kunitski2018}. 
The laser pulse is then turned on and the system is 
assumed to be imaged via COLTRIMS after a delay time~\cite{kunitski2018,coltrims}. 
In this technique, an
extremely short and intense probe
pulse, which ``rips off'' one electron of
each of the helium atoms, is applied and the ions are imaged. 
Since the probe pulse, to a very good approximation,
instantaneously projects the helium atoms to one particular 
configuration,
we do not simulate the imaging part of the experiment.
Repeated experimental measurements for the same time delay provide 
access to the quantum mechanical 
density distribution of the wave packet.
In what follows, 
the delay time 
is defined such that it is zero
when
the stretched Gaussian pulse first reaches its maximum.
Our calculations scan 
the delay time
from zero to many times 
$t_{\text{hold}}$.

We monitor
the correlator or alignment $C_{2}(r,t)$,
\begin{eqnarray}
C_2(r,t) = 
\frac{
\int_0^{\pi} \Psi^*(r,\theta,t) \cos^2 \theta \Psi(r,\theta,t) \sin \theta d \theta}
{\int_0^{\pi} |\Psi(r,\theta,t)|^2 \sin \theta d\theta}.
\end{eqnarray}
If $\Psi$ was independent of $\theta$
(as it is in the absence of the laser pulse), $C_2(r,t)$ would be
equal to $1/3$. Deviations from $1/3$ provide a
measure of the angle dependence that is introduced 
to the wave packet by the
laser pulse.
Importantly, after the laser is ``off'', i.e., 
after its intensity has decayed 
to a sufficiently small value, the coupling between different $l$ channels
vanishes and the populations of the 
different $l$ channels are independent of time.
The wave packet itself, however, continues to change
with time since the spatially-dependent
phases of the different partial wave components
continue to evolve. These phase factors imprint an $r$-dependent
interference pattern,
which varies with time
(see also Ref.~\cite{kunitski2018}), onto
the correlator $C_2(r,t)$.

\begin{figure*}
\centering
\includegraphics[angle=0, scale=0.5]{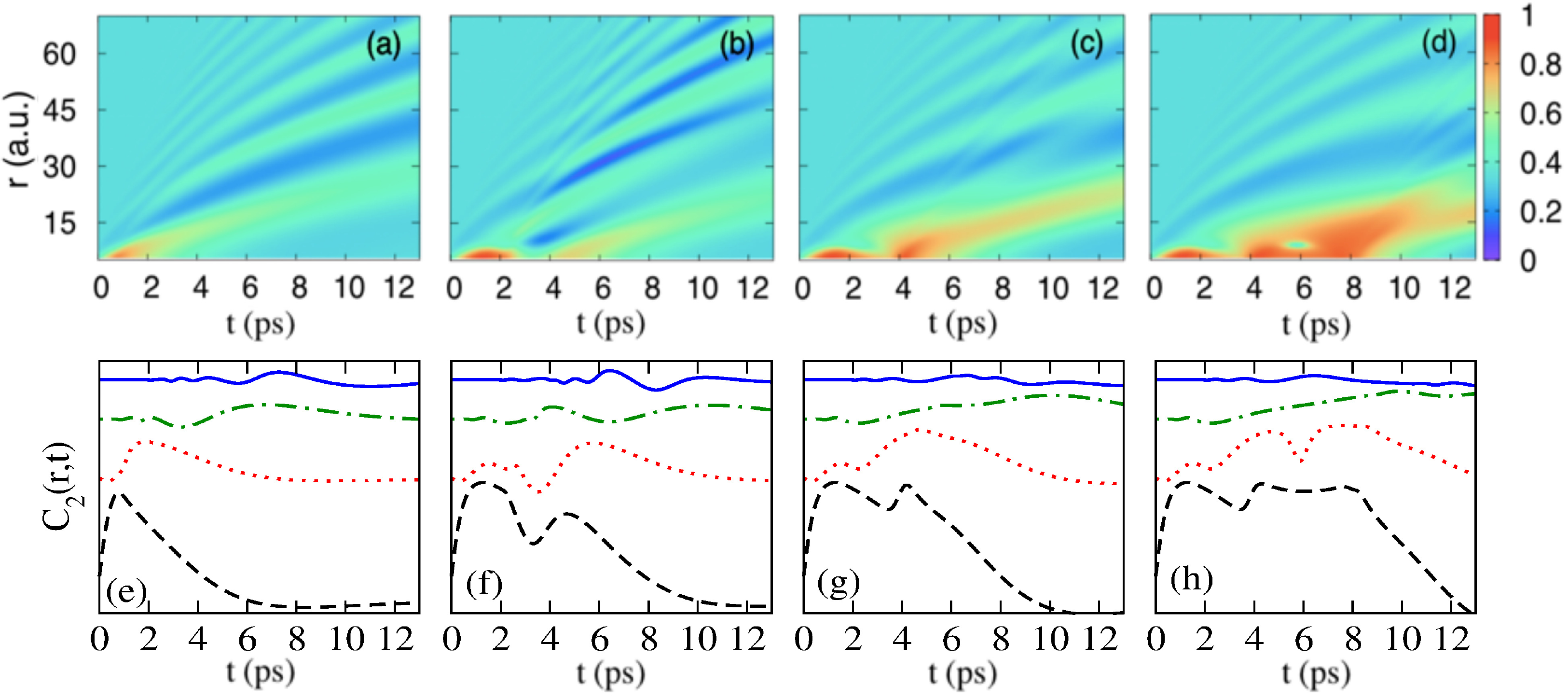}
\vspace*{0in}
\caption{(Color online)
Alignment signal for stretched Gaussian laser pulse
with $\epsilon_{0,\text{G}}=0.0843949a.u.$ 
and $\tau=311$~fs.
Results are shown for four different hold times:
(a) and (e) $t_{\text{hold}}=0.5$~ps,
(b) and (f) $t_{\text{hold}}=2$~ps,
(c) and (g) $t_{\text{hold}}=4$~ps,
and
(d) and (h) $t_{\text{hold}}=8$~ps.
Panels~(a)-(d)
show contour plots of the alignment signal
$C_2(r,t)$. A spherically symmetric wave packet
would yield an alignment signal of $1/3$.
The dashed, dotted, dash-dotted, and solid
lines in panels (e)-(h) show cuts of
$C_2(r,t)$ for $r=3\AA=5.669a.u.$, $r=5\AA=9.449a.u.$, $r=10\AA=18.90a.u.$, 
and $r=20\AA=37.79a.u.$,
respectively.
The curves are offset from each other
for ease of readibility.
}
\label{fig_dynamics1}
\end{figure*}

The upper row of
Fig.~\ref{fig_dynamics1} shows contour plots of $C_2(r,t)$
for fixed $\tau$ and $\epsilon_{0,\text{G}}$,
$\tau=311$fs and $\epsilon_{0,\text{G}} \approx 0.0844a.u.$ 
(intensity of $2.5 \times 10^{14}$W/cm$^2$),
and four different hold times,
i.e., for $t_{\text{hold}}=0.5$ps, $2$ps, $4$ps, and $8$ps.
The lines in the lower row show cuts, from bottom to top, for $r=3 \AA=5.669a.u.$,
$r=5 \AA=9.449a.u.$, $r=10\AA=18.90a.u.$, 
and $r=20 \AA=37.79a.u.$.
For the peak field strength used, 
the static system supports two bound states
that are dominated by the $s$-wave channel (see Figs.~\ref{fig_energy} 
and \ref{fig_ascatt}).

For the shortest $t_{\text{hold}}$ considered,
$C_2(r,t)$ is characterized by a fairly regular interference pattern,
whose maxima and minima move out with increasing time. 
Even though $t_{\text{hold}}$ is finite, the interference
pattern is quite similar to that observed and interpreted in a
very recent joint experiment-theory collaboration, which employed
an unstretched Gaussian pulse with 
$t_{\text{hold}}=0$,
the same $\tau$, and comparable field strength~\cite{kunitski2018}.
The pattern of the alignment signal 
can be traced back to the interference between 
the dissociating $l=2$ wave packet portion,
which gets populated as a consequence of the laser-molecule interaction,
and the broad spherically-symmetric background
portion (recall, the initial state is a pure $s$-wave state). 
Close inspection of $C_2(r,t)$ in the $t=0.5$~ps to $1$~ps window,
however, reveals that
the interference pattern is due to two
dissociating wave packet portions, one that is emitted starting at $t=0$
and another that is emitted for $t \gtrsim t_{\text{hold}}$.
This behavior becomes more prominent for larger $t_{\text{hold}}$
(see below).

The small $r$ behavior changes distinctly when $t_{\text{hold}}$ increases.
Figures~\ref{fig_dynamics1}(b)-\ref{fig_dynamics1}(d) display
oscillations of $C_2(r,t)$ at small $r$
[see also the dashed lines in
Figs.~\ref{fig_dynamics1}(f)-\ref{fig_dynamics1}(h)].
These oscillations, which are most prominent for the largest
hold time considered [Fig.~\ref{fig_dynamics1}(h)], are
roughly governed by the 
binding energy of the deepest-lying, $s$-wave
dominated transient state
that is supported by the static Hamiltonian with field strength
$\epsilon_{0,\text{G}}$. 
Its binding energy translates to 
about $3.800$~ps.
The time scale associated with the energy difference between the 
two $s$-wave dominated transient bound states is
equal to about $3.891$~ps, which is 
very 
close to the time scale set
by the binding energy
of the deep-lying transient state.
Indeed, we attribute the small-$r$ oscillations of $C_2(r,t)$
to two processes,
namely the interference between the 
wave packet portions corresponding to the two transient bound states
and the interference between the 
wave packet portions corresponding to the
deep-lying transient bound state
and unbound scattering states.
These 
interference
processes both contribute
to the population transfer between the $l=0$ and $l=2$ channels
and thus lead to oscillations in the alignment 
$C_2(r,t)$.

The
oscillations 
of $C_2(r,t)$
are reminiscent of revival dynamics
in rigid-rotor like molecules due to population transfer between 
different rotational states.
There are, however, important differences.
First, unlike for rigid-rotor molecules where multiple eigen energies
with spacings set by the rotational constant $B$ 
exist in the absence of the field, the deep-lying state that sets the time scale
in the helium dimer 
system
is transient.
Second, the 
$r$-dependence of the alignment $C_2(r,t)$,
as highlighted by the ``outgoing finger structure'' in Fig.~\ref{fig_dynamics1},
is unique to the non-rigid helium dimer. For rigid-rotor molecules,
this structure is absent.
Third, the broadness of the initial wave packet combined with the
fact that the laser-molecule interaction is dominant at small $r$
implies that only a small fraction of the wave packet gets 
``promoted'' to finite $l$ states.

As already alluded to above,
Figs.~\ref{fig_dynamics1}(b)-\ref{fig_dynamics1}(d)
show that the decay of the pump pulse
from strength $\epsilon_{0,\text{G}}$ to zero (this occurs for times just
a bit larger than $t_{\text{hold}}$)
triggers the ``emission'' of a second dissociating wave packet portion,
which can be attributed to the fact that
the population of the deep-lying transient bound state is no longer bound
when the laser intensity is negligible.
The second dissociating wave packet produces a new set of 
outgoing fingers
that are delayed by 
$t_{\text{hold}}$ compared to the first 
set of fingers and that ``collide''
with the first set of fingers.
The interference of the 
delayed outgoing wave packet portion with the first dissociating
wave packet portion leads, as can be seen nicely in
the $r=20 \AA$ cuts [solid lines in 
Figs.~\ref{fig_dynamics1}(f)-\ref{fig_dynamics1}(h)], to ``distorsions''
of the interference pattern.
In particular, it can be seen that $C_2(r,t)$
displays a regularly changing wave pattern for $t \le t_{\text{hold}}$
that changes notably for $t$ just a bit larger than $t_{\text{hold}}$.
For $t$ quite a bit larger than $t_{\text{hold}}$, $C_2(r,t)$
again displays a regularly changing wave pattern.

\begin{figure*}
\centering
\includegraphics[angle=0, scale=0.5]{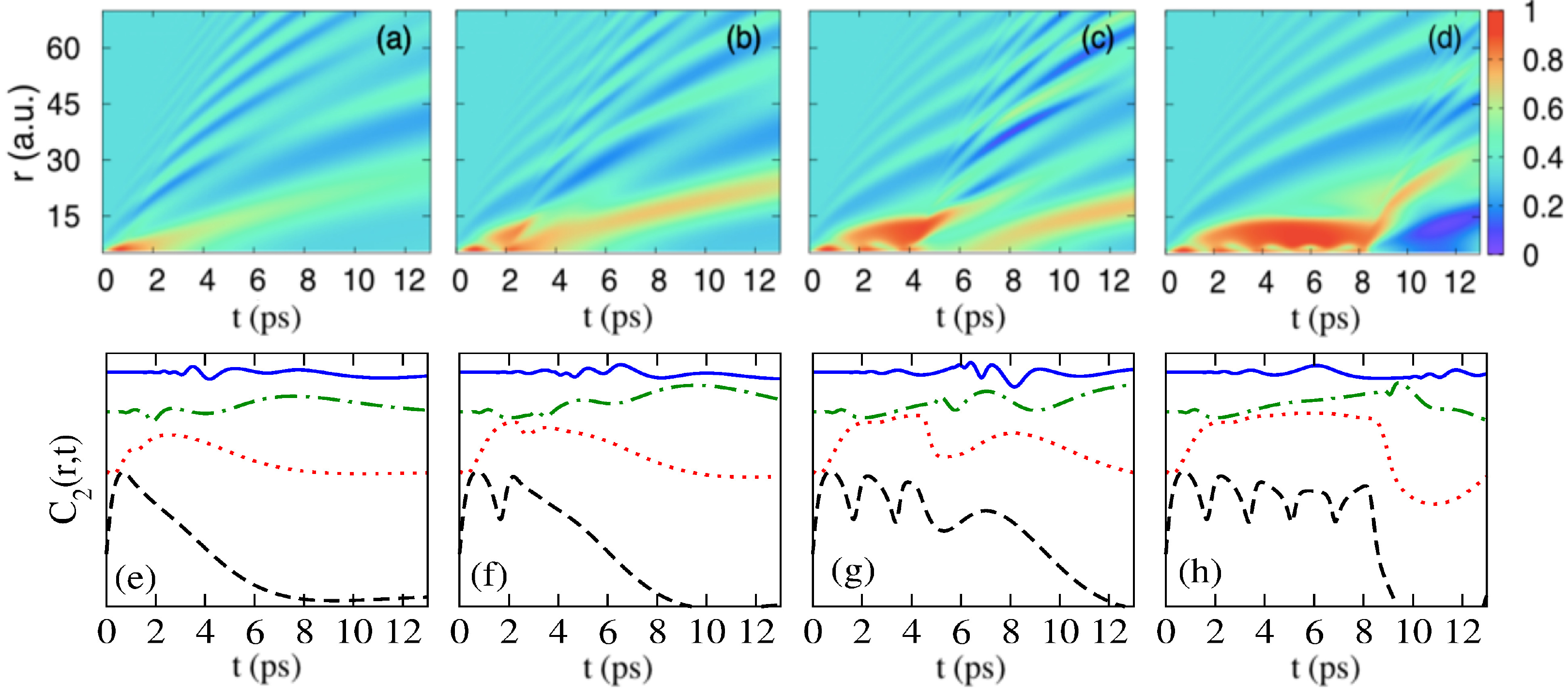}
\vspace{0cm}
\caption{(Color online)
Same as Fig.~\protect\ref{fig_dynamics1} but for 
a larger peak field strength, namely for
$\epsilon_{0,\text{G}}=0.11a.u.$.}
\label{fig_dynamics2}
\end{figure*}

Figure~\ref{fig_dynamics2} shows the same 
quantities as Fig.~\ref{fig_dynamics1}
but for 
a larger peak field strength, namely
for $\epsilon_{0,\text{G}}=0.11a.u.$. For
this field strength, the static $^4$He-$^4$He system
supports three $s$-wave dominated 
bound states and one $d$-wave dominated bound state.
The binding energy of the most strongly-bound transient state
translates to a time scale of $1.522$~ps.
Indeed, the small-$r$ region of the alignment signal
displays close to regular oscillations at roughly
this time scale. We do not expect perfect ``single-frequency'' 
oscillations since 
several transient eigen frequencies are expected to
contribute to the observed oscillatory pattern.
As in the weaker field strength case,
the emission of a second dissociating wave packet portion at times just a bit larger
than $t_{\text{hold}}$ is clearly visible in the 
alignment signal.
Comparison of Figs.~\ref{fig_dynamics1} and \ref{fig_dynamics2}
shows that the larger field strength has two primary effects. First, it leads
to a shortening of the oscillation period of the small-$r$ 
portion of $C_2(r,t)$. 
Second,  it enhances the contrast
of $C_2(r,t)$.
Besides these two effects, the overall behavior 
of $C_2(r,t)$ is quite similar.

Figures~\ref{fig_dynamics1} and \ref{fig_dynamics2}
demonstrate that pump-probe experiments on the $^4$He-$^4$He system
should provide evidence for the tunability of the 
bound state spectrum by an external electric field. However,
the alignment signal does unfortunately not---or if so 
rather indirectly---provide
access to
the number of field-induced bound states since the energy level spacing 
of the field-induced bound states 
is highly non-linear, leading to vastly different
time scales
governing the
interference
between the more weakly bound states.
Moreover, the highly non-linear spacing also makes it
difficult to distinguish between
oscillations 
in the alignment $C_2(r,t)$
due to the interference 
of wave packet portions corresponding to
the deepest-lying 
transient 
state
and
the most
weakly-bound 
transient
state and
oscillations
in the alignment $C_2(r,t)$
due to the interference
of wave packet portions corresponding to
the deepest-lying 
transient 
state
and
the 
transient 
scattering continuum.
The latter process contributes
also for peak field strengths $\epsilon_{0,\text{G}}$ that are smaller than 
$0.0715a.u.$,
i.e., for peak strengths
for which 
the static field Hamiltonian supports only one bound state. However, in
this field strength regime, the large time scale associated with
the
small binding 
energy makes the unambiguous experimental observation 
that the energy of the transient bound state has been tuned essentially impossible.

\section{Conclusion}
\label{sec_conclusion}
This work investigated static and 
dynamic
properties of 
helium-helium systems in the presence of an external electric
field.
All
three possible combinations of the two isotopes
$^3$He and $^4$He were investigated,
namely the
$^4$He-$^4$He,
$^3$He-$^4$He, and
$^3$He-$^3$He systems.
In the absence of an external field, only 
the $^4$He-$^4$He system supports a weakly-bound state (and 
only one).
When a static  external electric field is applied,
all three helium-helium systems display field-induced scattering resonances,
which are accompanied by the pulling-in of new two-body bound states.
The resonances and their characteristics were analyzed carefully.

Applying a stretched Gaussian laser pulse, the work investigated the 
signatures imprinted on the dynamics by the field-induced resonances.
For this analysis, we focused on the $^4$He-$^4$He system.
Assuming that the system is prepared in its only
bound state in the absence of an external field, the time evolution
during and after the stretched Gaussian pump pulse was investigated.
It was found that the time-evolving wave packet 
carries fingerprints of the field-induced bound states, in addition to
displaying 
dissociative dynamics that is associated with the fact that the pump
laser leads to the population of scattering states with zero and finite angular momenta.
It was commented that the experimental realization of the simulated 
scenarios is technically demanding but not impossible.

The response of diatomic rigid rotor-like molecules to intense 
laser pulses has been studied extensively in
the literature, both experimentally and theoretically.
The present dynamical study differs from these earlier works in that the
$^4$He-$^4$He system supports only
a single extremely weakly-bound state in the absence
of an external field. Thus, the notion of a rotor-like
spectrum does not
apply. As a consequence, the external field leads to a strong
coupling of the vibrational and rotational degrees of
freedom, with the populations 
of finite $l$ states dissociating.

\section{Acknowledgement}
\label{acknowledgement}
We are very grateful to 
R. D\"orner and M. Kunitski 
for extensive discussions,
which motivated and inspired this work. 
We are also very grateful to D. Fedorov for communication
related to Ref.~\cite{nielsen1999}.
Support by the National Science Foundation through
grant number
PHY-1806259
is 
gratefully acknowledged.
This work used
the OU
Supercomputing Center for Education and Research
(OSCER) at the University of Oklahoma (OU).

\appendix

\section{Coupling Matrix Elements $W_{l,l'}$}
\label{appendix_coupling}
To determine explicit
expressions for the coupling matrix elements $W_{l,l'}(r)$,
we rewrite
the laser-molecule interaction
$V_{\text{lm}}(r,\theta,t)$ as
\begin{eqnarray}
V_{\text{lm}}(r,\theta,t)=
g(t) \left[
\alpha_{0,0}(r) Y_{0,0} + \alpha_{2,0}(r) Y_{2,0}(\cos \theta) \right],
\end{eqnarray}
where 
\begin{eqnarray}
g(t) = -\frac{|\epsilon(t)|^2}{2},
\end{eqnarray}
\begin{eqnarray}
\alpha_{0,0}(r)=
\frac{\sqrt{4 \pi}}{3} \left[ \alpha_{\parallel}(r) + 2 \alpha_{\perp}(r) \right],
\end{eqnarray}
and
\begin{eqnarray}
\alpha_{2,0}(r)= 
\frac{\sqrt{16 \pi}}{3 \sqrt{5}} \left[ \alpha_{\parallel}(r) -
\alpha_{\perp}(r) \right].
\end{eqnarray}
Using this notation, $W_{l,l'}(r)$
becomes
\begin{eqnarray}
\label{eq_coupling}
W_{l,l'}(r)=
\alpha_{0,0}(r) \langle Y_{l,0}|Y_{0,0}|Y_{l',0}\rangle
+ \nonumber \\
\alpha_{2,0}(r) \langle Y_{l,0}|Y_{2,0}|Y_{l',0}\rangle
,
\end{eqnarray}
where the notation $\langle \cdot \rangle$ indicates an
integration over the angular degrees of freedom.


\begin{thebibliography}{100}

\bibitem{schoellkopf1994}
W. Sch\"ollkopf and J. P. Toennies,
Nondestructive Mass Selection of Small van der Waals Clusters,
Science {\bf{266}}, 1345 (1994).

\bibitem{tang1995}
K. T. Tang, J. P. Toennies, and C. L. Yiu,
Accurate Analytical He-He van der Waals 
Potential Based on Perturbation Theory,
Phys. Rev. Lett. {\bf{74}}, 1546 (1995).

\bibitem{janzen1995}
A. R. Janzen and R. A. Aziz,
Modern He-He potentials: Another look at binding energy, effective range 
theory, retardation, and Efimov states,
J. Chem. Phys. {\bf{103}}, 9626 (1995).

\bibitem{luo1996}
F. Luo, C. F. Giese, and W. R. Gentry,
Direct measurement of the size of the helium dimer,
J. Chem. Phys. {\bf{104}}, 1151 (1996).

\bibitem{schoellkopf1996}
W. Sch\"ollkopf and J. P. Toennies,
The nondestructive detection of the helium dimer and trimer,
J. Chem. Phys. {\bf{104}}, 1155 (1996).

\bibitem{grisenti2000}
  R. E. Grisenti, W. Sch\"ollkopf, J. P. Toennies, G. C. Hegerfeldt,
  T. K\"ohler, and M. Stoll,
Determination of the bond length and binding energy of
the helium dimer by diffraction from a transmission grating,
Phys. Rev. Lett. {\bf{85}}, 2284 (2000).

\bibitem{zeller2016}
S. Zeller, M. Kunitski, J. Voigtsberger, 
A. Kalinin, A. Schottelius, C. Schober, M. Waitz, 
H. Sann, A. Hartung, T. Bauer, M. Pitzer, F. Trinter, 
C. Goihl, C. Janke, M. Richter, G. Kastirke, M. Weller, 
A. Czasch, M. Kitzler, M. Braune, R. E. Grisenti, W. Sch\"ollkopf, 
L. Ph. H. Schmidt, M. S. Sch\"offler, J. B. Williams, 
T. Jahnke, and R. D\"orner,
Imaging the He$_2$ quantum halo state using a free electron laser,
PNAS {\bf{113}}, 14651 (2016).

\bibitem{chin2010}
C. Chin, R. Grimm, P. Julienne, and E. Tiesinga,
Feshbach resonances in ultracold gases,
Rev. Mod. Phys. {\bf{82}}, 1225 (2010).

\bibitem{nielsen1999}
E. Nielsen, D. V. Fedorov, and A. S. Jensen,
Efimov States in External Fields,
Phys. Rev. Lett. {\bf{82}}, 2844 (1999).


\bibitem{efimov70}
V. Efimov,
Energy levels
  arising from resonant two-body forces in a three-body system,
Phys. Lett. B {\bf{33}}, 563 (1970).

\bibitem{braaten2006}
E. Braaten and H.-W. Hammer,
Universality
  in few-body systems with large scattering length,
Phys. Rep. {\bf{428}}, 259 (2006).

\bibitem{naidon2016}
P. Naidon and S. Endo,
Efimov Physics: a review,
Rep. Prog. Phys. {\bf{80}}, 056001 (2017).


\bibitem{QiWei}
Q. Wei, S. Kais, T. Yasuike, and D. Herschbach,
Pendular alignment and strong chemical binding are
induced in helium dimer molecules by intense laser fields,
PNAS {\bf{115}}, E9058 (2018).

\bibitem{kunitski2018}
M. Kunitski, Q. Guan, H. Maschkiwitz, J. Hahnenbruch,
S. Echart, S. Zeller, A. Kalinin, M. Sch\"offler, L. Ph. H. Schmidt, T. Jahnke,
D. Blume, and R. D\"orner,
unpublished (2018).

\bibitem{friedrich1998}
B. Friedrich, M. Gupta, and D. Herschbach, 
Probing Weakly-Bound Species with Nonresonant Light: 
Dissociation of He$_2$ Induced by Rotational Hybridization,
Collect. Czech. Chem. Commun. {\bf{63}}, 1089 (1998).

\bibitem{becker}
A. Spott, A. Jaron-Becker, and A. Becker,
Time-dependent susceptibility of a helium atom in intense laser pulses,
Phys. Rev. A {\bf{96}}, 053404 (2017).

\bibitem{cencek2012}
  W. Cencek, M. Przybytek, J. Komasa, J. B. Mehl, B. Jeziorski, and
  K. Szalewicz,
  Effects of adiabatic, relativistic, and quantum electrodynamics
  interactions on the pair potential and thermophysical properties of
  helium,
J. Chem. Phys. {\bf{136}}, 224303 (2012).

\bibitem{friedrich1995}
B. Friedrich and D. Herschbach,
Alignment and Trapping of Molecules in Intense Laser Fields,
Phys. Rev. Lett. {\bf{74}}, 4623 (1995).

\bibitem{buckingham1973}
  A. D. Buckingham  and R. S. Watts,
  The polarizability of a pair of helium atoms,
  Mol. Phys. {\bf{26}}, 7 (1973).



\bibitem{cencek2011}
W. Cencek, J. Komasa, and K. Szalewicz,
Collision-induced dipole polarizability of 
helium dimer from explicitly correlated calculations
J. Chem. Phys. {\bf{135}}, 014301 (2011).

\bibitem{weiner1999}
J. Weiner, V. S. Bagnato, S. Zilio, and P. S. Julienne,
Experiments and theory in cold and ultracold collisions,
Rev. Mod. Phys. {\bf{71}}, 1 (1999).

\bibitem{marinescu1998}
  M. Marinescu and L. You,
  Controlling Atom-Atom Interaction at Ultralow
  Temperatures by dc Electric Fields,
Phys. Rev. Lett. {\bf{81}}, 4596 (1998).

\bibitem{yi2001}
  S. Yi and L. You,
  Trapped condensates of atoms with dipole interactions,
Phys. Rev. A {\bf{63}}, 053607 (2001).

  \bibitem{deb2001}
    B. Deb and L. You,
    Low-energy atomic collision with dipole interactions,
Phys. Rev. A {\bf{64}}, 022717 (2001).

\bibitem{newton_book}
  R. G. Newton,
  Scattering Theory of Waves and Particles,
  Dover Publications, Inc., Mineola, New York, 2nd Ed. (2002).


\bibitem{ticknor2005}
  C. Ticknor and J. L. Bohn,
  Long-range scattering resonances in strong-field-seeking
  states of polar molecules,
Phys. Rev. A {\bf{72}}, 032717 (2005).

\bibitem{kanjilal2008}
K. Kanjilal and D. Blume,
Low-energy resonances and bound states of aligned bosonic
and fermionic dipoles,
Phys. Rev. A {\bf{78}}, 040703(R) (2008).



\bibitem{johnson}
  B. R. Johnson,
The multichannel log-derivative method for scattering calculations,
J. Comp. Phys. {\bf{13}}, 445 (1973).

\bibitem{kosloff}
  T. H. Ezer and R. Kosloff,
  An accurate and effcient scheme for propagating the time dependent Schr\"odinger equation,
  J. Chem. Phys. {\bf{81}}, 3967 (1984).

\bibitem{kanjilal2007}
K. Kanjilal, J. L. Bohn, and D. Blume,
Pseudopotential treatment of two aligned dipoles
under external harmonic confinement,
Phys. Rev. A {\bf{75}}, 052705 (2007).

\bibitem{footnote}
Note that the $s$-wave scattering length defined in Ref.~\cite{nielsen1999} differs by a minus sign from our definition. 

\bibitem{bruch2000}
L. W. Bruch,
Electric field effects on the helium dimer,
J. Chem. Phys. {\bf{112}}, 9773 (2000).



\bibitem{aziz1991}
R. A. Aziz and M. J. Slaman,
An examination of ab initio results for 
the helium potential energy curve, 
J. Chem. Phys. {\bf{94}}, 8047 (1991).


\bibitem{RMP_alignment}
H. Stapelfeldt and T. Seideman,
{\em{Colloquium}}: Aligning molecules with strong laser pulses,
Rev. Mod. Phys. {\bf{75}}, 543 (2003).

\bibitem{review_lemeshko}
M. Lemeshko, R. V. Krems, J. M. Doyle, and S. Kais,
Manipulation of molecules with electromagnetic fields,
Mol. Phys. {\bf{111}}, 1648 (2013).

\bibitem{coltrims}
J. Ullrich, R. Moshammer, R. D\"orner, O. Jagutzki, V. Mergel, H. Schmidt-B\"ocking, 
and L. Spielberger,
Recoil-ion momentum spectroscopy,
J. Phys. B {\bf{30}}, 2917 (1997).


\end{thebibliography}
\end{document}